# Epistemic and Aleatoric Uncertainty Quantification in Weather and Climate Models

Laura A. Mansfield*: laura.mansfield@physics.ox.ac.uk, Atmospheric, Oceanic and Planetary Physics, University of Oxford, Oxford, United Kingdom,

Hannah M. Christensen: hannah.christensen@physics.ox.ac.uk, Atmospheric, Oceanic and Planetary Physics, University of Oxford, Oxford, United Kingdom,

*Corresponding author.

## Abstract

Representing and quantifying uncertainty in physical parameterisations is a central challenge in weather and climate modelling, and approaches are often developed separately for different timescales. Here, we introduce a unified framework for analysing uncertainty in parameterisations across weather and climate regimes. Using the Lorenz 1996 system as a testbed for simplified chaotic dynamics, we quantify uncertainties in a subgrid-scale parameterisation using a Bayesian Neural Network (BNN). This allows us to disentangle aleatoric uncertainty, arising from internal variability in the training data, and epistemic uncertainties, arising from poorly constrained parameters during training. At runtime, we sample uncertainties in line with stochastic approaches in weather models and perturbed-parameter methods in climate models. On weather timescales, aleatoric uncertainty dominates, underscoring the value of stochastic parameterisations. On longer, climate timescales and under changing forcings, accounting for both types of uncertainty is necessary for well-calibrated ensembles, with epistemic uncertainty widening the range of explored climate states, and aleatoric uncertainty promoting transitions between them. Constraining parameter uncertainty with short simulations reduces epistemic uncertainty and improves long-term model behaviour under perturbed forcings. This framework links concepts from machine learning with traditional uncertainty quantification in Earth system modelling, offering a pathway toward seamless treatment of uncertainty in weather and climate prediction.

# 1. Introduction

*a. General Circulation Models*

General circulation models (GCMs) simulate the Earth's atmosphere, ocean, land surface, and sea ice. They consist of a dynamical core which solves the governing equations of fluid motion on a 3D grid for the Earth, and physical parameterisations or closures, that represent unresolved subgrid-scale processes, such as radiation, convection, clouds and aerosol microphysics, atmospheric gravity waves, ocean eddies, and land surface processes (Christensen & Zanna, 2022) . Parameterisations are typically based on simplified equations or empirical relationships and make several assumptions which can introduce a large source of uncertainty into GCM output.

Recently, machine learning (ML) and artificial intelligence (AI) approaches have gained popularity for learning subgrid-scale parameterisations (e.g., Behrens et al., 2022; Christensen & Zanna, 2022; Heuer et al., 2024; Rasp et al., 2018; Souza et al., 2020; Ukkonen & Chantry, 2025; Yu et al., 2024; Yuval & O'Gorman, 2023). However, while uncertainty quantification (UQ) has long been a crucial aspect of parameterisation development (e.g., Berner et al., 2017), it is often overlooked in ML-based approaches (Christensen et al., 2024). This paper explores UQ methods that could be leveraged as we transition towards AI-enhanced GCMs. We examine the roles of epistemic (model) and aleatoric (data) uncertainty and how they fit in with the traditional view of uncertainties in GCMs. We advocate that UQ should remain an essential part of parameterisation development, regardless of whether it is physics-based or data-driven.

*b. Uncertainties in General Circulation Models*

Understanding and communicating uncertainties are an essential part of weather and climate prediction (Gigerenzer et al., 2005). Meteorological agencies, governments, energy providers, the agriculture sector, insurance companies, among others need to make decisions based on GCM output. For instance, public weather forecasts rely on probabilistic predictions (Gneiting & Katzfuss, 2014). These are especially important for low-probability but high-impact events, such as heavy rainfall and flooding (Cloke & Pappenberger, 2009). These may require advanced warnings or even government action to reduce the impact on society. On longer timescales, GCMs are used to generate projections of future climate change projections are essential for policy-makers developing long-term strategies for climate adaptation (e.g., Calvin et al., 2023; Stainforth et al., 2005). Furthermore, scientific studies that predict climate model response to different forcings, physics, or geography, must also consider potential uncertainties before making conclusions (e.g., Forster et al., 2013; Murphy et al., 2004).

1) WEATHER FORECASTING

On weather prediction timescales, uncertainty in the initial conditions become amplified due to the chaotic processes in the atmosphere. This is **initial condition uncertainty** and can be captured by perturbing the input state and running ensembles (Slingo & Palmer, 2011). There are also uncertainties associated with the model formulation, particularly in the choices and assumptions made when developing parameterisations. This is known as **model uncertainty** (Slingo & Palmer, 2011). Usually, this term refers to uncertainties in the **structure** and **parameters** of the model or parameterisation. When developing parameterisations, there is also the issue that for a given resolved state, there can be many possible unresolved states. This is the **subgrid variability** and can be dealt with through **stochastic parameterisations** (Berner et al., 2017).

**Deterministic parameterisations** assume that, given resolved state variables, the tendencies from unresolved processes can be estimated with a deterministic function derived from the mean grid-box behaviour. This assumption is valid when there is a clear scale separation between resolved and unresolved processes (Berner et al., 2017; Christensen & Zanna, 2022; Palmer, 2019). However, as model resolution increases and more processes become partially resolved, this scale separation breaks down. In these "grey zone" regimes, the processes are not explicitly resolved but too few subgrid-scale processes exist within a single grid-cell to define the mean grid-box behaviour. For instance, this occurs at around the order of 10s to 100s km for atmospheric organised moist convection (Christensen & Zanna, 2022). Instead of taking the average over many subgrid-scale processes, we must sample a single realisation of that subgrid-scale process, creating stochastic parameterisations (Berner et al., 2017). An example of implementing this in practice is the stochastically perturbed parameterisation tendency (SPPT) scheme, which combines a deterministic estimate of the most likely subgrid-scale tendency with a stochastic perturbation to represent its variability (often correlated in time and space, Buizza et al., 1999).

2) CLIMATE CHANGE PROJECTIONS

In contrast, on climate timescales we are generally interested in *climate change projections*, where we predict the climate response to a forcing. This is a boundary condition problem rather than an initial condition problem. Initial condition uncertainty is not considered a large source of uncertainty on these timescales (Slingo & Palmer, 2011), as we are interested in climate statistics. Furthermore, subgrid variability may be less pronounced due to the larger scale separation. Instead, the main three types of uncertainty are **internal variability** (which causes natural variations on decadal timescales), **model uncertainty** (in model structure or parameters, which define how different models respond differently to the same forcing) and **scenario uncertainty** (our lack of knowledge about the future forcing itself) (Hawkins & Sutton, 2009). Internal variability can be captured by repeating climate forcing experiments with different initialisations for the ocean state to capture variations in the response that occurs on decadal timescales. In contrast, the latter two types are more relevant on longer, centennial timescales. Here, we will not consider scenario uncertainty, which requires running simulations under several possible future pathways, and instead focus on model uncertainty.

Similarly to NWP, climate model uncertainty arises from the different modelling choices that can be made during model development. One approach to represent model uncertainty is to consider multi-model ensembles from many different modelling centres (e.g., multi-model CMIP simulations (Eyring et al., 2016)), which captures uncertainty regarding the model structure and parameterisation assumptions (**structural uncertainty** (Rougier, 2007)). Within a given model, there are also uncertainties around the model parameters that define parameterisations, which we call **parametric uncertainty**. These can be quantified by running "Perturbed Parameter Ensembles" (or "Perturbed Physics Ensembles", PPEs) which involve sampling model parameters according to domain expertise and running ensembles of simulations (e.g., Christensen et al., 2015b; Eidhammer et al., 2024; Murphy et al., 2004, 2007; Sengupta et al., 2021; Stainforth et al., 2005).

*c. Reducing Uncertainties in GCMs*

While uncertainties can never be fully eliminated, parametric uncertainties can be reduced in a process known as **calibration**, a key aspect of UQ and model development of both weather and climate models (e.g., Carslaw et al., 2013; Dunbar et al., 2021; Sengupta et al., 2021; Souza et al., 2020; Williamson et al., 2017). Calibration focuses on reducing parametric uncertainty, by constraining parameters based on past observations. Techniques that leverage PPEs are often used to eliminate parameter values that produce model output inconsistent with observations, through history matching (Couvreux et al., 2021; King et al., 2024; Raoult et al., 2024; Williamson et al., 2013), approximate Bayesian computation (Watson-Parris et al., 2021), or ensemble Kalman methods (Dunbar et al., 2021; Mansfield & Sheshadri, 2022). Many of these methods are aided by machine learning emulators such as Gaussian process emulators, which reduce the number of expensive GCM integrations required. In contrast to this, initial condition uncertainty, subgrid variability and internal variability are internal properties of the Earth system and our observations of it and therefore cannot be reduced through model development.

*d. The rise in ML parameterisations*

Over the last few years, we have witnessed a rise in ML-based parameterisations, which are trained on existing physics-based parameterisations (e.g., Chantry et al., 2021; Espinosa et al., 2022; Ukkonen, 2022), cloud-resolving models embedded within GCM grid-cells (e.g., Hu et al., 2025; Rasp et al., 2018; Yu et al., 2024), coarse-grained high-resolution GCMs simulations (e.g., Giles et al., 2024; Grundner et al., 2022; Henn et al., 2024; Heuer et al., 2024; Morcrette et al., 2025; Ross et al., 2023; Watt-Meyer et al., 2024; Yuval & O'Gorman, 2023), or observations (e.g, Miller et al., 2025). There has been substantial interest in stochastic ML-based parameterisations, which includes stochasticity in a similar manner to SPPT described above (Christensen et al., 2024). Stochastic ML schemes are primarily used to improve model skill (e.g., Gagne II et al., 2020; Giles et al., 2024; Guillaumin & Zanna, 2021; Nadiga et al., 2022; Perezhogin et al., 2023) but some studies also focus on their use for UQ (e.g., Behrens et al., 2025; Mansfield & Sheshadri, 2022; Miller et al., 2025). Here, we argue that while AI is becoming increasingly used in weather and climate modelling, evaluating and quantifying uncertainty remains critical for ensuring trust and credibility in predictions (Haynes et al., 2023; McGovern et al., 2022). As we demonstrate here, probabilistic ML methods, such as Bayesian deep learning, also provide a natural framework for uncertainty quantification.

*e. This study*

In this study, we quantify uncertainties in a subgrid parameterisation, decomposing them by source and across different timescales. We use the Lorenz, 1996 (L96) model as a case study to demonstrate how UQ should be carried out and to compare different approaches to sampling uncertainties. We use L96 because it simulates both large and small-scale variables and their interactions and exhibits chaotic properties similar to that of the real atmosphere but can be run at a significantly lower computational cost than a full GCM. This makes it a suitable testbed for parametrisations. It has been used to test stochastic parameterisations (Arnold et al., 2013; Wilks, 2005) and machine learning parameterisations (Chattopadhyay et al., 2020; Gagne II et al., 2020; Parthipan et al., 2023; Rasp, 2020) for numerical weather prediction. It has also been used to explore the uncertainties associated with parameterisations on climate timescales, for instance, in PPE studies (Christensen et al., 2015a). Although a simplified model, we can use it to draw analogies to more complex GCMs, while considering a range of prediction timescales.

In Section 2, we discuss the framework of uncertainties commonly used in the ML community that describe aleatoric uncertainty, coming from the data, and epistemic uncertainty coming from the model, and how these fit in with the traditional uncertainty viewpoint used in the weather and climate communities. In Section 3, we outline model and methods used, including the Lorenz 1996 model and how a deterministic neural network parameterisation can be used in place and how we quantify uncertainties using Bayesian Neural Networks, including an assessment of uncertainties in an "offline" setting (i.e., data pre-generated). In Section 4, we analyse the uncertainties once the parameterisations are coupled back in the Lorenz 1996 model, which we refer to as "online". We consider these on weather forecasting timescales and in Section 5, we consider these on a climate timescale. Finally, Section 6 draws conclusions and discusses how we can assess uncertainties in weather and climate models that use ML parameterisations going forward.

## 2. Background

*a. Types of Uncertainties in Machine Learning*

Here, we approach uncertainty quantification from a machine learning perspective, where we categorise uncertainty into two types: **aleatoric uncertainty** and **epistemic uncertainty** (Hüllermeier & Waegeman, 2021).

The term "aleatoric" comes from the Latin word *alea*, meaning "game of chance". Aleatoric uncertainty is used to describe the variability in a system that is due to inherently random effects (Haynes et al., 2023; Hüllermeier & Waegeman, 2021). It represents the statistical or stochastic nature of a system, such as flipping a coin or rolling a dice and this type of uncertainty cannot be reduced. In the ML literature, aleatoric uncertainty refers to uncertainty in the data. This could include data generated from a stochastic process, but it can also include data originating from unobserved variables, which means there is no longer a one-to-one mapping from inputs to outputs.

The term "epistemic" comes from the Greek word *epistēmē*, which means knowledge or understanding. Epistemic uncertainty is caused by a lack of knowledge about the best model for a system. This can be thought of a systematic uncertainty and can be reduced with more knowledge or understanding of the system which can come from more data. In the ML literature, epistemic uncertainty refers to uncertainty in the ML model. This includes uncertainty in model structure (i.e., model architecture and number of layers and neurons), and uncertainties in model parameters (i.e., weights and biases).

*b. Aleatoric and Epistemic Uncertainties in GCMs*

We can relate these types of uncertainties to those used by the Earth system modelling community. Initial condition uncertainty and internal variability are both forms of aleatoric uncertainty, as they arise from the chaotic nature of the Earth system. Here, we will also consider subgrid variability as aleatoric uncertainty because it arises from the internal variability that can occur within a grid-box. From an ML perspective, we can also view this as uncertainty in the training data that arises when knowledge of the large-scale state, $X$, does not uniquely define the subgrid-tendency, $U$. This type of uncertainty is irreducible given the task is to learn the $U$ from $X$ alone. In contrast, we will consider structural and parametric uncertainties as epistemic uncertainty, because they are modelling choices that we make during model development. We also consider scenario uncertainty to be a form of epistemic uncertainty, as it comes from lack of knowledge in the forcing that should be applied. These types of uncertainties, the timescales they dominate on, and the typical approach used to estimate them are summarised in Table 1.

Table 1. Types of uncertainties in GCMs on varying timescales, whether they are aleatoric or epistemic, and the traditional approach used in weather and climate

| Timescale | Type of Uncertainty in GCMs | Aleatoric | Epistemic | Typical Approach |
|---|---|---|---|---|

| Weather | Initial Condition | x | | Perturbed IC ensemble (atmosphere, derived from observations) |
|---|---|---|---|---|
| Weather | Subgrid Variability (informs parameterisations, i.e., training data) | x | | Stochastic Parameterisation |
| Seasonal to Decadal | Internal Variability | x | | Perturbed IC ensemble (for longer timescales, perturbed sea surface temperatures) |
| All | Structural uncertainty | | x | Multi-model ensemble |
| Climate | Parametric uncertainty | | x | Perturbed parameter ensemble |
| Climate | Scenario uncertainty | | x | Multi-scenario ensemble |

*b. Aleatoric and Epistemic Uncertainties in GCMs that use ML parameterisations*

ML is becoming increasingly used for parameterisations in GCMs. We expect ML-based parameterisations to have similar associated uncertainties to physics-based schemes. In the ML framework, the goal is to learn the relationship between the large-scale variables and the subgrid variables from the training data. The subgrid variability creates a noisy dataset and can be viewed as the uncertainty in the training data, in other words, the aleatoric uncertainty.

Epistemic uncertainties also exist within ML-based parameterisations that are analogous to the types of uncertainties discussed previously for conventional parameterisations. While physics-based parameterisations typically only contain a handful of parameters that lead to parametric uncertainty, ML-based parameterisations have a large number of parameters to learn. This means the parametric uncertainties and the relationships between them are likely more complex. Structural uncertainties are also present for ML-based parameterisations, since there are choices to make regarding ML algorithms, architectures, and number of parameters.

Furthermore, the use of ML gives rise to another form of epistemic uncertainty that isn't traditionally considered in UQ: "*out of regime*" or "*out of sample*" uncertainty. This occurs when an ML algorithm is trained on data that is generated by one distribution but is applied to a dataset that was generated by a different distribution. ML algorithms are known to perform poorly during extrapolation, especially if they have a large number of parameters and/or have been overfit. This could be an issue for climate modelling, where training data may come from present day climate, but the ML parameterisation could be applied under a different future climate scenario.

In this study, we quantify epistemic and aleatoric uncertainties in an ML parameterisation. Note, we do not consider structural error and instead, assume a fixed neural network structure for our parameterisation. We use Bayesian Neural Networks (BNNs) to capture both epistemic uncertainty, which represents parametric uncertainty and out-of-regime uncertainty, and aleatoric uncertainty, which represents the sub-grid variability. BNNs provides us a with a way to clearly distinguish between these two forms of uncertainty.

## 3. Methods

*a. The Lorenz 1996 Model*

We adopt the two-layer L96 system as a simplified model of chaotic dynamics to explore UQ methods for ML parameterisations (Lorenz, 2006). It can be viewed as toy model for mid-latitude atmospheric dynamics around a latitude circle. In the one-layer version, the variables are $X_k$ where $k = 1, \ldots, K$, with periodic boundary conditions, i.e., $X_{K+1} = X_1$. The variables evolve following

$$\frac{dX_k}{dt} = -X_{k-1}(X_{k-2} - X_{k+1}) - X_k + F$$

Equation 1

where $F$ is an external forcing. This can be solved with a choice of numerical methods, such as the Euler method.

The two-layer version extends this to include large-scale variables, $X$, and small-scale variables, $Y$, defined from $j = 1, ... J$, between each of the large-scale variables Fig. 1.

$$\frac{dX_k}{dt} = -X_{k-1}(X_{k-2} - X_{k+1}) - X_k + F + \frac{-hc}{b} \sum_{j=J(k-1)+1}^{kJ} Y_j$$

Equation 2

$$\frac{dY_j}{dt} = -cbY_{j+1}(Y_{j+2} - Y_{j-1}) - cY_j + \frac{hc}{b} X_{int[j-1/J]+1}$$

Equation 3

where the int notation refers to the integer value of the term inside the brackets (i.e., the closest $X$ variable) and $b$, $c$, and $h$ are user-defined parameters of the system that describe the spatial scale ratio, the temporal scale ratio and a coupling constant, respectively. Following (Arnold et al., 2013; Wilks, 2005), we use $K = 8$, $J = 32$, $b = 10$, $c = 10$, $h = 1$ and $F = 20$. The term highlighted in red is the **subgrid-scale term** that couples the small-scale variables to the large-scale variables.

Since the two-layer system must solve for the small-scale variables, it usually requires more advanced numerical methods such as fourth order Runge-Kutta scheme (RK4), used here with a timestep of $\Delta t = 0.001$. This motivates the task of replacing the small-scale variables with a parameterisation, allowing a coarser timestep and cheaper numerical scheme to be used.

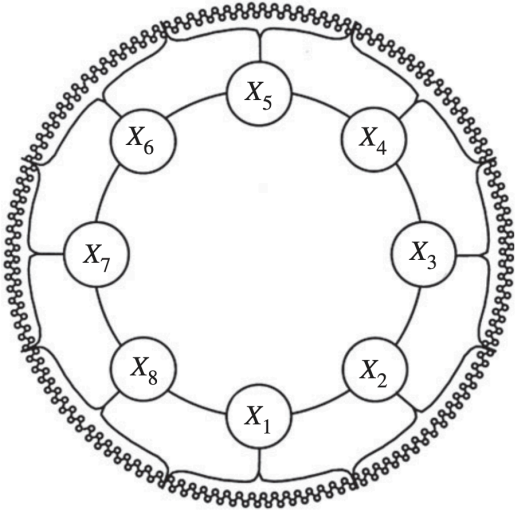

Fig. 1 Schematic showing the Two Layer L96 set-up with $K = 8$ large-scale variables, $X$, and $J = 32$ small-scale variables, $Y$, coupled to each $X$ variable and cyclic boundary conditions. Reproduced from Wilks (2005).

To reduce computational cost, we can replace the coupled system of equations with a single equation that describes only the large-scale variables with an additional parameterisation term, $U_k = f(X_k)$, as follows:

$$\frac{dX_k}{dt} = -X_{k-1}(X_{k-2} - X_{k+1}) - X_k + F + {\color{red}f(X_k)}$$

Equation 4

This is the 'forecast model'. This parameterisation term aims to captures the small scale variables, $Y$, without explicitly evaluating Equation 2. The function $f$ can be learned from a training dataset. This can then be solved with a cheaper numerical method such as RK2, and with a coarser forecast timestep, $\Delta t_f = 0.005 = 5\Delta t$, reducing computational burden by around 5 ×, assuming the function $f$ is cheap to evaluate.

b. *Training a Neural Network Parameterisation*

To learn the function, $f$, we can train a neural network. Training data can be generated by simulating the two-layer coupled system and to use these $X_k$ terms to learn $U_k = f(X_k)$ in the forecast model (Equation 4).. Rather than directly storing the subgrid-scale term ($\frac{-hc}{b}\sum_{j=J(k-1)+1}^{kJ} Y_j$ in Equation 1), as done in Balwada et al. (2024) and Rasp (2020), we estimate the subgrid-scale term from the $X_k$, as done in Arnold et al. (2013), Gagne II et al. (2020), Parthipan et al. (2023), and Wilks (2005). Given a dataset $X_k$ from the two-layer model, stored at intervals $\Delta t_f$, we can rearrange Equation 4 to obtain

$$[U_k]_t = \frac{[X_k]_{t+\Delta t_f} - [X_k]_t}{\Delta t_f} - [-X_{k-1}(X_{k-2} - X_{k+1}) - X_k + F]_t$$

Equation 5

This is representative of how parameterisations are trained on high-resolution datasets, where the sub-grid tendencies are not available and instead must be estimated through carrying out a coarse-graining in space (e.g., Heuer et al., 2024; Morcrette et al., 2025; Watt-Meyer et al., 2024; Yuval & O'Gorman, 2023). ML parameterisations typically use only variables from a single vertical column, since information from neighbouring grid cells is unavailable. Consistent with this approach, our goal is to predict $U_k$ as a function of $X_k$ alone. We will treat all $K$ points on the circle as separate data points. For the training dataset, we select 100 independent data points from a timeseries over a span of $T = 1000$ MTU, equivalent to around 1½ "atmospheric years", estimated by considering the error doubling time in the system (Wilks, 2005). Including the spatial dimension with $K = 8$, this gives 800 samples for training, shown in Fig. 2. Other ML parameterisation studies typically use $O(10^6 - 10^7)$ samples to train networks with $O(10^5 - 10^6)$ parameters (e.g., Grundner et al., 2022; Ukkonen & Chantry, 2025; Yu et al., 2024), giving comparable sample-to-parameter ratios.

We start with a basic deterministic fully connected neural network with two hidden layers, each with 16 nodes (similar to Rasp, 2020). We write a neural network as

$$U = f_\theta(X)$$

where $X$ and the inputs, $U$ are the outputs and $\theta$ are the network weights. We train the network to minimise mean squared error. The black line in Fig. 2 shows the resulting neural network fit. This neural network does pick up the general trend, but it is evident that there is noise within the data coming from the subgrid variability that cannot be captured by the deterministic NN. In the following section, we train a BNN to capture this as a form of aleatoric uncertainty, as well as quantifying epistemic uncertainties from uncertainty in the model parameters.

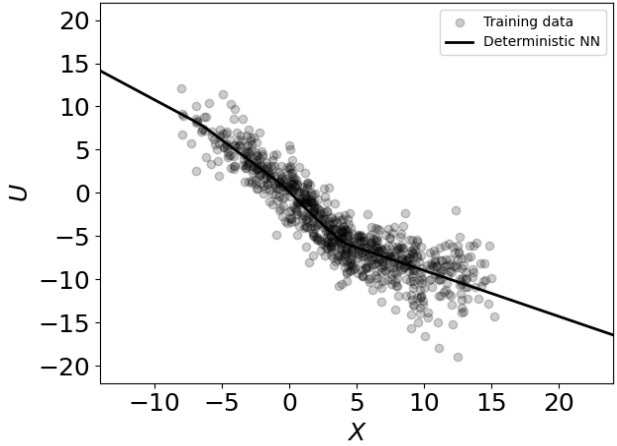

Fig. 2 Training data where *x*-axis shows the large-scale variables (inputs to neural network) and the *y*-axis shows the subgrid tendency (outputs of neural network). The black solid line shows the prediction from the deterministic neural network.

*c. Learning Uncertainties using a Bayesian Neural Network*

To quantify and separate the sources of uncertainty into epistemic and aleatoric, we use Bayesian Neural Networks (BNNs). BNNs extend traditional neural networks with a Bayesian treatment of the neural network parameters (the weights and biases, $\theta$). We use this approach because in recent years, neural networks have become the primary choice of machine learning parameterisations (e.g., Hu et al., 2025; Ukkonen & Chantry, 2025; Yuval & O'Gorman, 2023), due to their flexibility, simplicity to implement, and widespread use in the machine learning community. We will write a BNN as

$$U = f(X|\theta)$$

Instead of learning fixed weights, BNNs treat the weights as probability distributions and use Bayes' theorem to update these distributions based on observed data (Goan & Fookes, 2020). This captures uncertainty in the parameters or *parametric uncertainty*, a form of epistemic uncertainty. We can also treat the model output as a probability distribution, allowing us to capture inherent noise in the dataset, or *aleatoric uncertainty*. Since this approach is probabilistic, the approach to training differs from standard neural network training. Rather than seeking *parameters* that best fit the data, $\theta$, we seek *probability distributions, $p(\theta|X, U)$* that are most likely to generate the dataset $(X, U)$. See

Supplementary Text S1 for a full description of how we learn these probability distributions using variational inference.

Fig. 3 shows a simplified schematic of the BNNs used here. The distributions on the connections between nodes highlight that all parameters (weights and biases) have associated probability distributions. Here, we will assume that the probability distribution over the parameters can be approximated with a multivariate Gaussian distribution, with a mean and covariance matrix, allowing for correlations between the parameters (Barber & Bishop, 1997; Hinton & van Camp, 1993).

We also represent the data with a Gaussian distribution, with mean $\mu$ and the variance $\sigma^2$. Fig. 3a shows that the output layer predicts this mean $\mu$ and we separately learn $\sigma^2$ as a fixed parameter across the entire dataset. This does not consider that some regions of the input space may be associated with increased sub-grid variability, which should be represented as higher aleatoric uncertainty. Fig. 2 above shows that this may indeed be case, as there is increased noise for larger values of $X$. Allowing the aleatoric uncertainty to vary with $X$ is known as **heteroscedasticity**, and is a common challenge in Bayesian machine learning (Kendall & Gal, 2017). We will consider a heteroscedastic version of the BNN by learning the variance on the output layer as a function of the inputs, shown in Fig. 3b. The BNN predicts two values as outputs instead of just one: the mean $\mu$ and the variance $\sigma^2$. This approach is more complicated because the aleatoric uncertainty now depends upon the parametric uncertainty within the BNN, meaning there is not such a clear separation between the two forms of uncertainty.

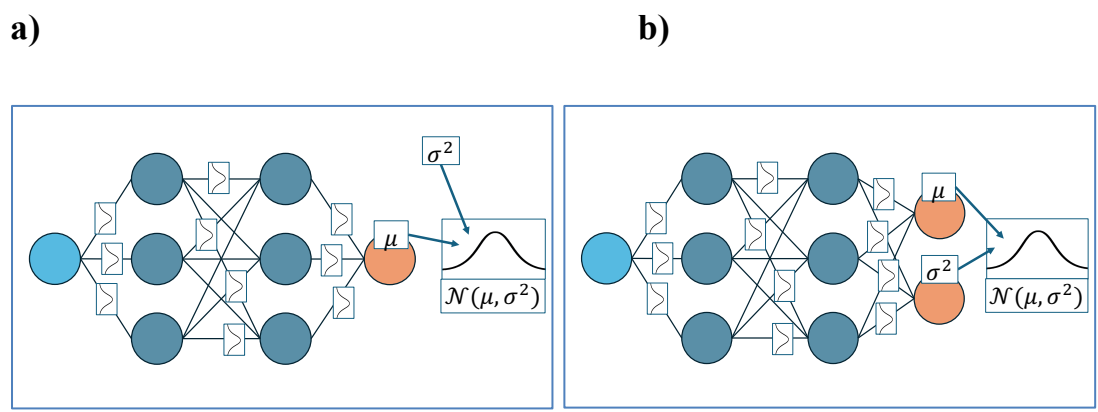

Fig. 3 Schematic of Bayesian Neural network. All network weights are assumed to have associated probability distributions, represented by the normal distributions on the nodes in this diagram (epistemic

uncertainty). The output layer is also assumed to be a normal distribution with mean $\mu$ and variance $\sigma^2$ which represents the noise in the data (aleatoric uncertainty). The BNN in a) learns $\sigma^2$ as a fixed scalar, independent of input values (known as "homoscedasticity") while in b) the BNN directly outputs $\sigma^2$, allowing it to depend on the inputs (known as "heteroscedasticity"). Note that this is a simplified representation of the BNN used here, as we use more nodes within each hidden layer and the weights are treated as multivariate normal distributions to capture correlations between the weights.

Once the BNN is trained, we can sample from the parameter distributions to estimate epistemic and aleatoric uncertainties. For a given value of $X$, a single pass through the network requires sampling the parameters $\theta$ from the variational distribution. To estimate probabilities requires repeatedly sampling parameters and passing the inputs through the network. We can sample *epistemic uncertainty* alone, by sampling the weights only and taking the mean prediction at the output layer. Alternatively, we can sample the *aleatoric uncertainty* alone by fixing the weights at their median values to predict $\mu$ (and $\sigma^2$ for the heteroscedastic version in Fig. 3b), then sampling from the output Gaussian distribution. By sampling both the weights and the outputs, we can sample *both* sources of uncertainty together.

*c. Offline Uncertainties*

We first assess the uncertainties in BNN predictions in an *offline* setting, meaning the input $X$ data has already been generated by the full two-layer system in Equations 1-2 (Bracco et al., 2025). Fig. 4 shows the BNN mean prediction across the domain and the shading shows 2 standard deviation uncertainty obtained when sampling from aleatoric, epistemic or both sources of uncertainty as described above. The points show the training data (same as Fig. 2). Increasing the size of the training dataset did not significantly reduce epistemic uncertainty (Supplementary Fig. 1). Fig. 4a shows the results of homoscedastic BNN that assumes fixed aleatoric uncertainty across the entire dataset (Fig. 3a), while Fig. 4b shows the heteroscedastic version where the BNN also predicts the variance as a function of the input data, allowing the aleatoric uncertainty to vary with $X$ (Fig. 3b). The latter appears to better capture the variations in the data. Both show epistemic uncertainty to be lower in the centre of the dataset where the parameters are more constrained but increase as towards the edges of the dataset where there is increased out-of-regime uncertainty. For the rest of the paper, we use heteroscedastic BNN because it appears to capture aleatoric uncertainty better, unless stated otherwise.

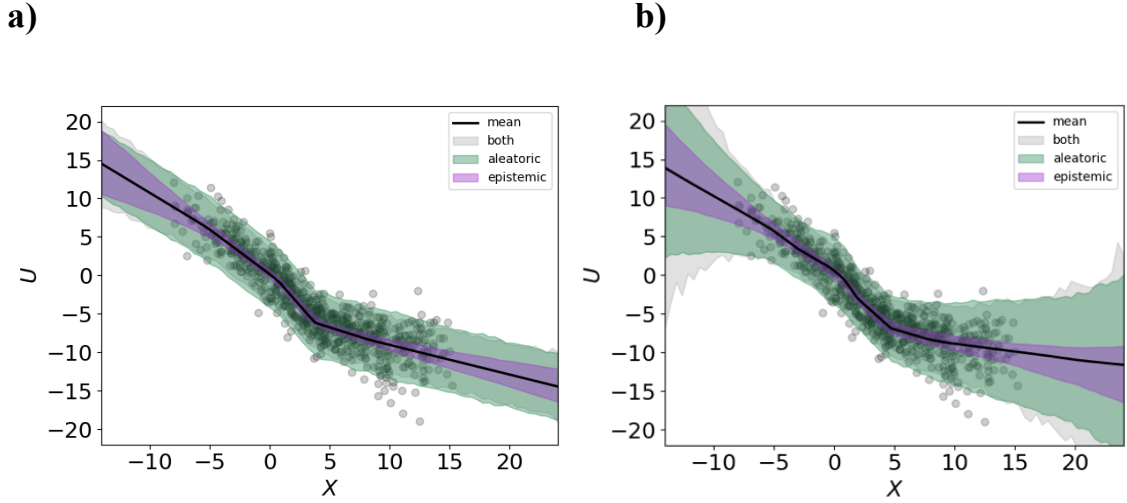

Fig. 4 Offline results for aleatoric and epistemic uncertainty, shading shows 2 std. away from mean. **a)** Aleatoric uncertainty is learned as a scalar parameter that is fixed across the entire dataset (homoscedastic), **b)** Aleatoric uncertainty varies across the dataset (heteroscedastic) by setting the BNN up to predict two outputs: a mean and a variance.

## 4. Online Coupling on Weather Timescales

After training, we couple the machine learning parameterisation back into the One Layer L96 through Equation 4 where $f(X)$ is the trained ML algorithm. This is used to update the state $X$ at the next timestep, which in turn is used to estimate $f(X)$ at the following timestep. This creates a feedback between the large-scale dynamics and ML prediction of the subgrid-scale physics. We call this *online* evaluation (Bracco et al., 2025). We will compare to a "truth" experiment that uses the Two Layer L96 model, initialised with the state at the end of the training dataset and run for $T = 1000$.

For the weather forecasting problem, we use the BNN to forecast trajectories starting from $N = 100$ initial conditions from the truth dataset, separated by $T = 10$ (which corresponds to ~50 atmospheric days, more than sufficient for the initial conditions to be uncorrelated).

a. *Independent noise at each timestep*

At each timestep, to estimate $U$ in Equation 4 with the BNN, $U = f(X|\theta)$ we must sample the weights $\theta$. This creates a stochastic parameterisation where we obtain a different outcome each time we estimate $U$ for a given $X$. To estimate uncertainties, we must run multiple ensemble members, here using $N = 50$. We run the BNN in three configurations: "aleatoric", where the model parameters are kept fixed at their median values but we sample from the output layer ($\mathcal{N}(\mu, \sigma^2)$ in Fig. 3b), "epistemic" where the model parameters are sampled and the output layer is deterministic ($\mu$ in Fig. 3b), and "both" where we sample from the model parameters and the output layer. Here, the sampling methods are entirely independent at each timestep (i.e., a white noise process). This means that, unlike most stochastic parameterisations (e.g., SPPT, Buizza et al., 1999), the random component of the subgrid-scale prediction is not correlated in time.

Fig. 5a,c,e shows the trajectories of one variable given one initial condition under each of these settings. We find that the ensemble members start to diverge from each other after $T \sim 0.6$. They diverge from the truth slightly earlier, after $T \sim 0.4$ indicating poor skill. The epistemic uncertainty is significantly lower than the aleatoric uncertainty and takes longer for the ensemble members to diverge. "Both" appears to follow the aleatoric uncertainty more closely.

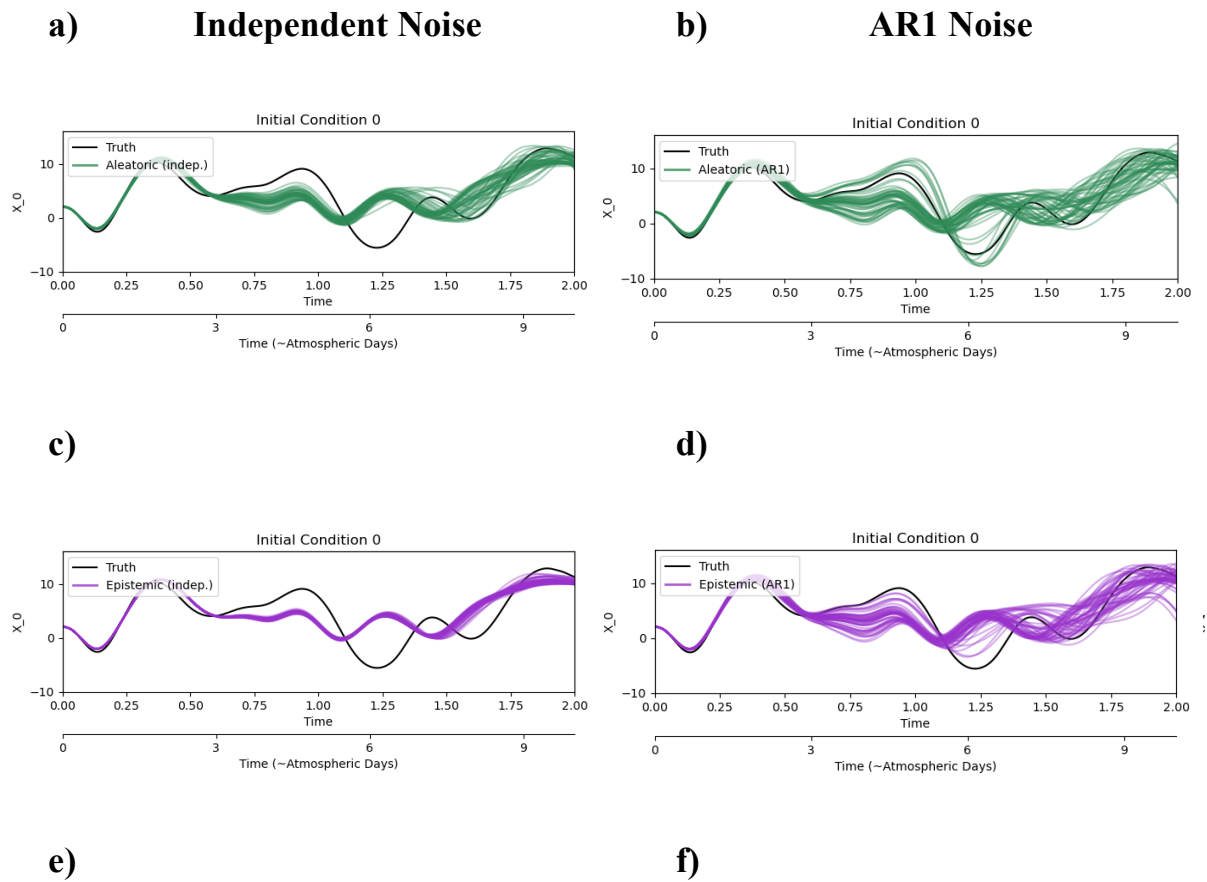

a) **Independent Noise**  b) **AR1 Noise**

c)  d)

e)  f)

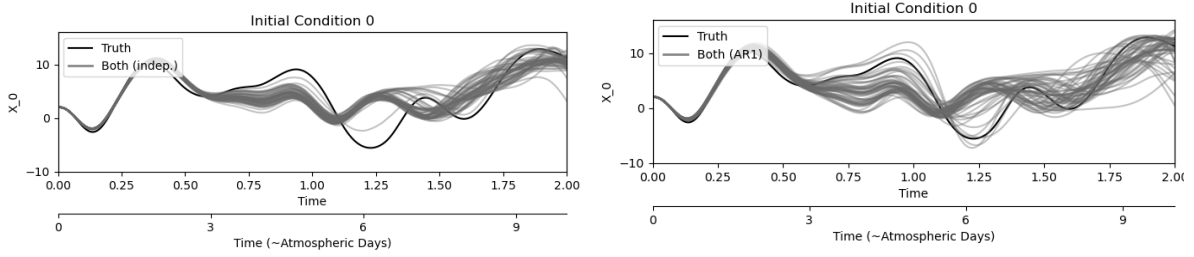

Fig. 5 Trajectories for variable $X_0$, generated by the one-layer L96 model with the stochastic parameterisation where each line represents a different ensemble member and the black line shows the true, two-layer L96 model (black line). The parameterisations sample (a-b) aleatoric uncertainty only, (c-d) epistemic uncertainty only and (e-f) both types of uncertainty. Panels on the left (a,c,e) show parameterisations that are sampled independently at each timestep, while panels on the right (b,d,f) use an AR1 process to include temporal correlations.

*b. Including temporal correlation*

Fig. 5a,c,e show the ensemble members diverge from the truth before they spread out from each other. Unlike many stochastic parameterisations used in operational weather forecasting models (Buizza et al., 1999; Shutts, 2005), there is no temporal correlation between $U_k$. Many studies have found that temporal correlation is essential in stochastic parameterisations to both improve skill and the spread of a forecast (Arnold et al., 2013; Berner et al., 2017). In Fig. 5b,d,f, we include temporal correlation following an Auto Regressive Order 1 (AR1) process, similar to the method used in Arnold et al., (2013).

This approach assumes that the stochastic parameterisation can be broken down into two parts: a deterministic component (here, the neural network with weights fixed at their median values) and a random noise component (Wilks, 2006). To maintain correlation with the previous timestep, the random component combines the noise from the previous timestep with a new noise term, known as the *innovation*. We use the autoregressive parameter, estimated from the lag-1 correlation (0.985) to enforce a suitable correlation between successive timesteps, giving a correlation timescale of 0.33 MTU. To define the innovation to be added at each timestep, we estimate the variance associated with the aleatoric or epistemic uncertainty for the input, $X_t$. Full details are provided in the Supplementary Text S2.

Fig. 5b,d,f show the trajectories that use the AR1 parameterisation for aleatoric, epistemic, and both sources of uncertainty, respectively. As before, simulations that sample

both types of uncertainty have similar spread to those that sample aleatoric uncertainty only. For all simulations, using an AR1 sampling approach leads ensemble members to diverge faster than with independent sampling. The ensemble members spread out enough to capture the truth over much of the simulation, showing a major improvement in the reliability of the forecast. Note that we see similar results when using the homoscedastic BNN, although adding the AR1 process increases aleatoric uncertainty more significantly (Supplementary Fig. 2).

We test how well the ensemble members agree with the truth more robustly by repeating simulations with multiple different initial conditions. Fig. 6 shows the overall RMSE and spread against time in solid and dashed lines respectively. These are computed as the square root of the sum of the squared error (RMSE) and variances (spread) over 100 simulations with different initial conditions. Fig. 6a shows that when we sample independent noise at each timestep, the spread is consistently smaller than the RMSE, indicating under-dispersive ensembles. When AR1 noise is introduced, Fig. 6b shows that the spread and error match well and grow at similar rates, indicating well-calibrated ensembles. The errors also grow more slowly. Note this is not the case for the homoscedastic ensembles, where including aleatoric uncertainty leads to an over-dispersive ensemble (Supplementary Fig. 3). In all cases, the spread in the aleatoric ensembles is almost identical to the spread in the ensembles that sample both forms of uncertainty. This suggests that they are not distinct, independent sources of uncertainty and that the contribution from epistemic uncertainty is small enough to ignore on these timescales. This confirms the validity of past studies on ML stochastic parameterisations for weather timescales, where only aleatoric uncertainty from the subgrid variability is captured by the ML scheme (Behrens et al., 2025; Guillaumin & Zanna, 2021; Perezhogin et al., 2023).

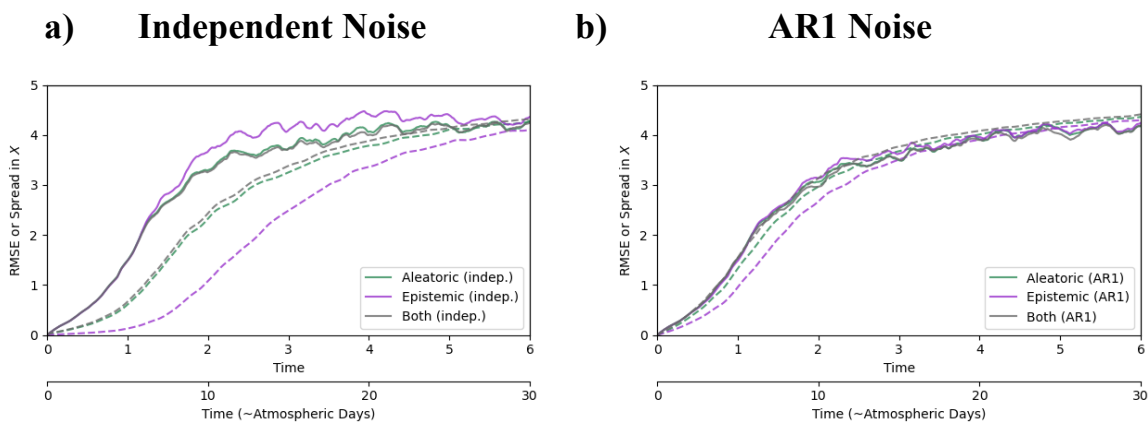

Fig. 6 RMSE (solid line) and spread (dashed line) both averaged over 100 different simulations for a) independent noise and b) AR1 noise.

*c. Reliability*

The **reliability** of a forecast refers to how well predicted probabilities align with observed probabilities (Arnold et al., 2013; Leutbecher, 2009; Leutbecher & Palmer, 2008). For ensemble forecasts, the ensemble spread should be an indication of the error in the ensemble mean. This is known as **statistical consistency**. We test for statistical consistency by considering spread against error for the 100 independent forecasts, evaluated at the same time ($t = 0.5$ MTU, approx. 2½ atmospheric days). Following Leutbecher (2009), we compute the ensemble mean and variance for each sample. We sort the samples by increasing variance and partition them into bins of size 100. Root Mean Squared (r.m.s.) spread is defined as the square root of the mean ensemble variance within each bin and r.m.s. error is defined as the square root of the variance of the ensemble mean error within each bin. Fig. 7 shows the r.m.s. spread on the *x*-axis against r.m.s. error on the *y*-axis for a) the independent noise simulations and b) the AR1 simulations. Points lying in the upper left triangle of the plot indicates under-dispersive ensembles, where spread < error, whereas the points lying in the lower right triangle are a feature of an over-dispersive ensemble, where spread > error. Points lying along the $y = x$ line indicate a well calibrated ensemble. Fig. 7a show all independent noise simulations fall within the upper left triangle, indicating under-dispersive ensembles, while Fig. 7b show that using the AR1 process increases the spread significantly. Although the epistemic ensembles are still underdispersive, when aleatoric uncertainty is included, the ensembles appear better calibrated. Most points falling along the $y = x$ line before $T = 0.5$ MTU, but as the simulations continue beyond $T = 0.5$ MTU, the r.m.s error increases more rapidly than the spread. Note that this differs from the homoscedastic BNN, where aleatoric uncertainty is constant for all input variables. Fig. 7d shows that including an AR1 process leads to over-dispersive aleatoric ensembles (r.m.s. spread is larger than r.m.s. error). This suggests that a heteroscedastic treatment of aleatoric uncertainty is important for well-calibrated ensembles.

**a) Independent Noise, Heteroscedastic BNN**    **b) AR1 Noise, Heteroscedastic BNN**

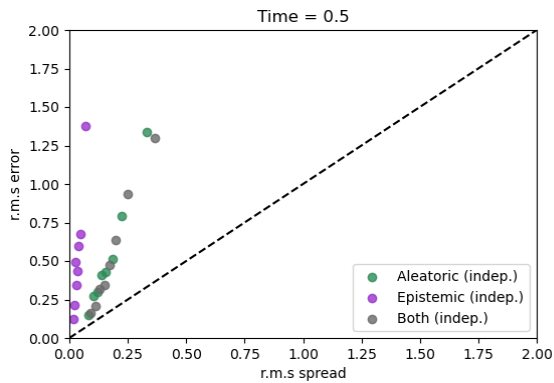
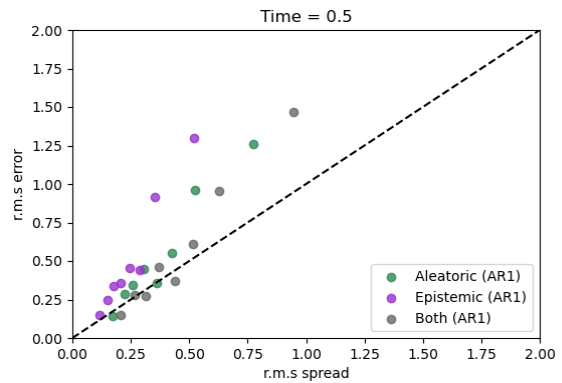

**c) Independent Noise, Homoscedastic BNN**    **d) AR1 Noise, Homoscedastic BNN**

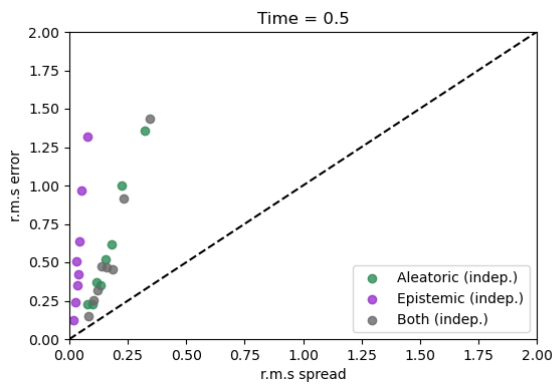
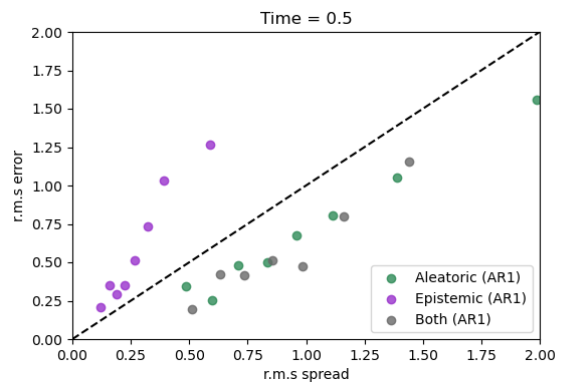

Fig. 7 r.m.s spread against r.m.s. error for a) independent noise simulations and b) AR1 simulations for the BNN that samples aleatoric (green), epistemic (purple) and both (grey) uncertainty. Each point represents a bin of 100 samples, over which the r.m.s. spread and r.m.s. error are calculated. The black dashed line shows the $y = x$ line, which represents a well-calibrated ensemble.

## 5. Online Coupling on Climate Timescales

*a. Climate simulations*

For climate prediction, we are interested in long-term statistics, so for this we run a long simulation with $T = 1000$, corresponding to around 14 years. Fig. 8a shows that over these timescales, the distributions of $X$ for all simulations match the truth closely but they all slightly underestimate the tails of the distribution, highlighted by the lower panel which shows the difference between the distribution and the true distributions. There are no major differences between any sampling approaches. The deterministic parameterisation performs well and behaves very similarly to the epistemic parameterisation, likely because the epistemic uncertainty introduced is small (especially when using AR1 process, which has a short correlation timescale (~0.33 MTU) and therefore its effect averages out over longer timescales). This highlights the need for more sophisticated diagnostics beyond long term probability distributions.

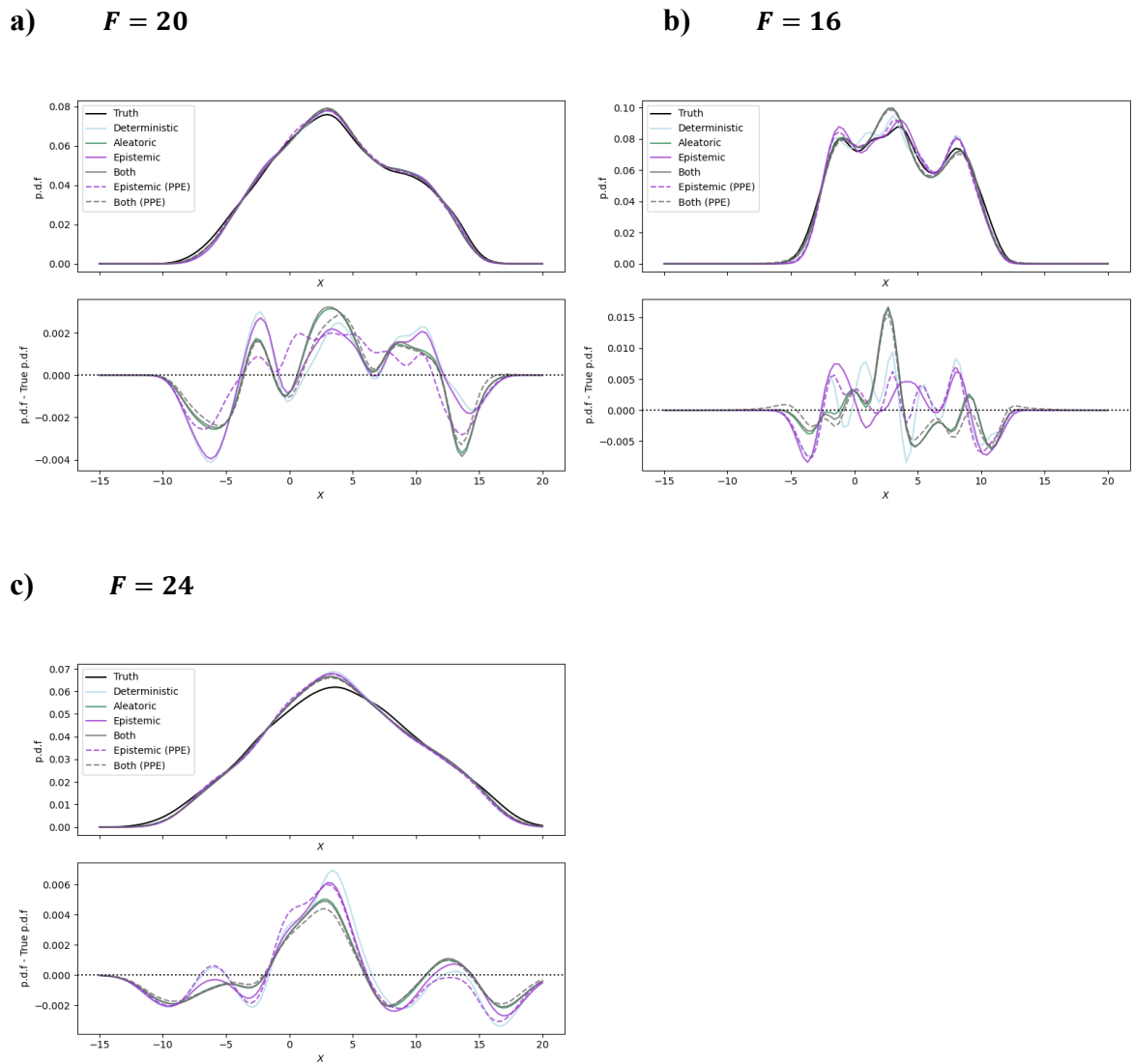

Fig. 8 Probability distribution functions (p.d.f) and the difference between the p.d.f and the True p.d.f over $X$ for longer integrations up to $T = 1000$ for a) the "baseline climate" with same setup as used for the training data with $F = 20$, and perturbed climates with b) $F = 16$, and c) $F = 24$.

*b. Climate change scenarios*

On climate timescales, GCM users are typically interested in climate *change* given a change in forcing or emissions scenario. Here, we simulate a climate change experiment by perturbing the forcing, $F$, in Equation 4. The baseline climate that is used to generate the training data is $F = 20$ for all cases. Fig. 8b shows the distribution over $X$ for a decrease in forcing, where $F = 16$ and Fig. 8c shows the distribution over $X$ for an increase in forcing where $F = 24$. We do not see significant improvements in the distributions when moving from a deterministic forecast to a stochastic one that includes uncertainties. This suggests that on these timescales, deterministic parameterisations may be sufficient.

*c. Perturbed Parameter Ensembles*

It is worth considering how epistemic uncertainty is usually treated by the climate modelling community. For a conventional parameterisation, parametric uncertainty is typically estimated through "perturbed parameter ensembles" (PPEs), where uncertainties on the parameters are defined based on domain knowledge. For each ensemble member, parameter values are sampled once and are fixed for the duration of the simulation. PPEs are used to quantify and, where possible, reduce parametric uncertainty by further constraining parameter values (Karmalkar et al., 2019; Murphy et al., 2007; Sexton et al., 2021). Here, we carry out a PPE by sampling the BNN weights once for each ensemble member and holding them fixed throughout the simulation. We expect this to be a more realistic representation of epistemic uncertainty, which should remain constant across timescales, rather than fluctuating on timescales associated with correlations in the subgrid residual (as in the AR1 based approach). These simulations are also shown in Fig. 8 by the purple dashed lines, but again, do not show consistent differences from the deterministic forecast. The grey dashed line shows simulations that combine a PPE with AR1 sampling for aleatoric uncertainty. These show slight improvements in capturing the tails of the distributions in the climate change simulations, but the differences are minor.

*d. Modes of variability*

When considering long-term climate, we are usually concerned with obtaining the correct modes of variability. We typically measure these using scaler metrics, which either measure the phase of an oscillation or characterise phenomena using long term statistics. Examples in the real world include the El Niño Southern Oscillation (ENSO) occurring on timescales of 3-5 years which is quantified by NINO3-4 indices (Wang et al., 2017); the North Atlantic Oscillation (NAO) which is quantified by the NAO index (Hurrell et al., 2003); and the Quasi-Biennial Oscillation (QBO) in stratospheric winds, which can exist in an easterly or westerly phase and can be characterised by its period or amplitude (Schenzinger et al., 2017).

Lorenz noticed that the L96 model exhibits oscillatory modes with similar properties to these climate modes of variability (Lorenz, 2006). The spatial patterns over $X$ over the circular domain can fall under two possible regimes: one which has a single wave (wavenumber $k = 1$) or one with two waves (wavenumber $k = 2$). Following the same method as (Christensen et al., 2015b), we identify these regimes using principal component analysis (PCA, or empirical orthogonal functions, EOFs). Applying PCA decomposition over the truth timeseries highlights four modes of variability which explain 78% of the variance. The first two modes show a wavenumber $k = 1$ pattern which dominates 38% of the time, while the second two modes show a wavenumber $k = 2$ pattern which dominates 62% of the time. The two modes for each pattern exist because they are out of phase. We will compare whether the neural network parameterisation simulations exhibit the same ratio of time spent in each regime.

We explore the PPE simulations here to see if parametric uncertainty can lead to systematic biases and different climate states. Here, we will consider a different climate state to be one that has different preference for either the $k = 1$ regime or the $k = 2$ regime. We run the simulations for T=1000 MTU, each with 50 ensemble members. We also repeat this for 10 different initial conditions, to reduce the influence of the starting point.

a)
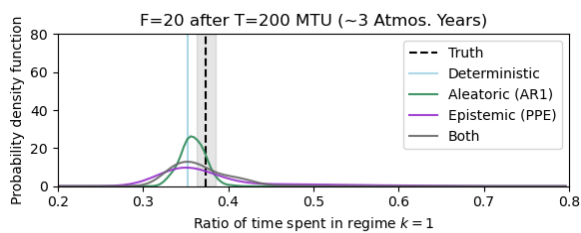

b)
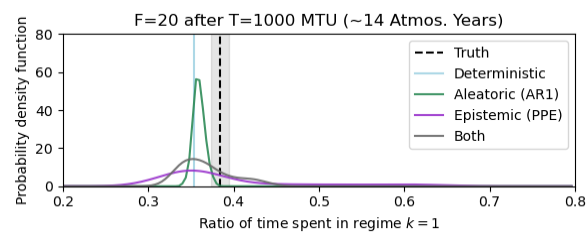

c)
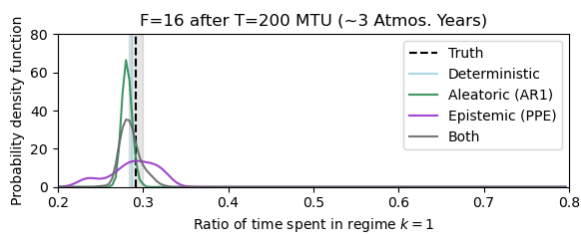

d)
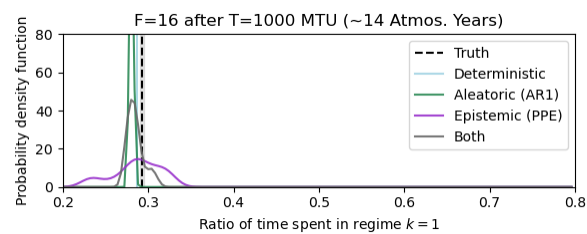

e)

f)

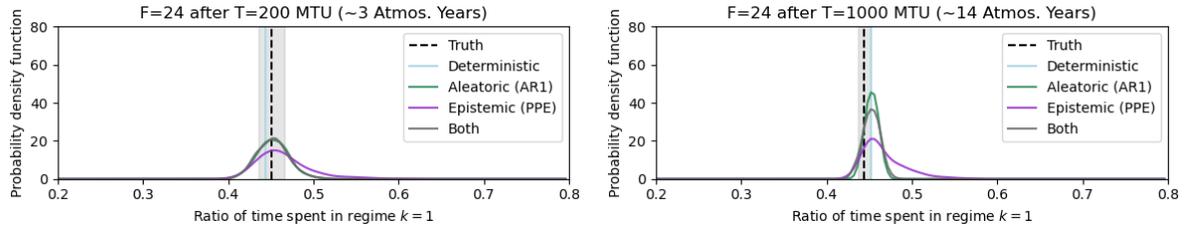

Fig. 9 Distributions showing ratio of time spent in regime $k = 1$ across all ensemble members for (a-b) $F = 20$, (c-d) $F = 16$ and (e-f) $F = 24$. The black dashed line shows the truth and the grey shading around this shows the standard deviation across a ten-member ensemble of true simulations initialised with different initial conditions. The light blue line shows the mean from the deterministic prediction. The purple, green and grey lines show the distributions across the epistemic, aleatoric and both ensembles, calculated using kernel density estimation across all 50 ensemble members and ten initial conditions. These are computed after (a,c,e) time T=200 MTU (~3 atmospheric years) and after (b,d,f) T=1000MTU (~14 atmospheric years).

Fig. 9 shows the distributions across all ensemble members in the ratio of time spent in the $k = 1$ regime. The black dashed line shows the true ratio of time spent in the $k = 1$ regime, with the grey shading showing the standard deviation in this value across the ten different initial conditions. When $F = 20$, Fig. 9a shows a mismatch between the deterministic prediction and the truth, making any uncertainty estimate beneficial over a deterministic approach. The aleatoric uncertainty is narrower than epistemic uncertainty after time T=200 (~3 atmospheric years). As the simulation continues out to T=1000 MTU (~14 atmospheric years), Fig. 9b shows that aleatoric uncertainty becomes even narrower, while the distribution representing the epistemic uncertainty remains similar. Here, the aleatoric simulations are overconfident, with the peak of the distribution lying to the left of the truth, indicating most ensemble members under-represent the time spent in the $k = 1$ regime.

This narrow aleatoric uncertainty arises because it reflects instantaneous subgrid variability which fluctuates rapidly, on timescales associated with the AR1 process (~0.33 MTU). These short-term fluctuations do not affect the long-term ratio of time spent in each regime. In contrast, fixed parameter perturbations derived from epistemic uncertainty can induce persistent shifts in regime behaviour (Christensen et al., 2015a). Some ensemble members remain in either the $k = 1$ or the $k = 2$ regime for extended periods. This leads to the large uncertainty that remains constant across the simulation. While this ensemble does capture the truth, the persistent regime behaviour is not realistic. The ensemble sampling both aleatoric and epistemic uncertainty shows that introducing stochastic variability from the aleatoric component can promote transitions between regimes, preventing members from remaining in a single state for too long. This shows that including short-term stochasticity can improve regime behaviour, consistent with results from GCM experiments (Dawson & Palmer, 2015). Together, these results highlight the importance of jointly considering both epistemic and aleatoric uncertainty.

Fig. 9c-f shows the distributions for the climate change experiments, with panels (c-d) corresponding to $F = 16$ and (e-f) to $F = 24$. Although the BNN was trained on $F = 20$, it successfully reproduces the direction of the shift in regime preference under different forcings. This generalisation is encouraging for climate change applications, assuming the behaviour holds in a full GCM. As before, the aleatoric ensemble is overconfident and fails to capture the truth, particularly for $F = 16$. Here, introducing parameter perturbations increases ensemble diversity enough to capture the truth, indicating the potential benefit of including epistemic uncertainty in climate change experiments.

### e. Reducing Model Uncertainty

A key part of UQ lies in reducing model uncertainty in a process known as **calibration**. Calibration aims to constrain parameter values based on past observations and therefore focuses on reducing epistemic uncertainty. Here, we highlight a simple approach to calibration that uses the PPE we have generated. From the $F = 20$ simulations, at T=200MTU (~3 atmospheric years), we select ensemble members that fall within 1 standard deviation of the truth (the grey shading in Fig. 9a). This gives us 6 ensemble members from the epistemic ensemble, or 10 from the ensemble sampling both uncertainty types. These parameter choices are better constrained and less likely to produce persistent regime behaviour. Fig. 10 shows the resulting constrained ensembles at T=1000 MTU (~14 atmospheric years) in bold for (a-b) $F = 20$, (c-d) $F = 16$, and (e-f) $F = 24$. For (a-b) $F = 20$, the constrained ensembles are centred on the truth and the spread is reduced significantly. This shows the potential to constrain parametric uncertainty once a parameterisation in coupled online, which could be particularly valuable for full GCMs. The constrained ensemble that samples both uncertainty types agrees particularly well with the truth (Fig. 10b). This suggests an additional benefit of including aleatoric uncertainty during calibration, as it captures subgrid variability and reduces regime persistence. If reproduced in full GCMs, this approach could improve calibration of physics-based and ML parameterisations (Hourdin et al., 2017; Williamson et al., 2017).

For the climate change experiments, the epistemic distributions (Fig. 10c for $F = 16$ and Fig. 10e for $F = 24$) move closer to the truth with reduced uncertainty, even though they have been constrained on the $F = 20$ simulations. This demonstrates potential for improving climate model parameterisations through online fine-tuning of parameters. For the ensemble that samples both uncertainty types, when $F = 16$ (Fig. 10d), the constrained ensemble slightly underestimates the truth, although it still provides an improvement upon the unconstrained ensembles, especially those that sample aleatoric uncertainty alone (Fig. 9d). We do not see major improvements in the reduced ensemble that samples both for $F = 24$ (Fig. 10f). Calibration based directly on the climate change simulations ($F = 16$, $F = 24$) is expected to improve performance and would be recommended for climate change applications in full GCMs, if available.

a)
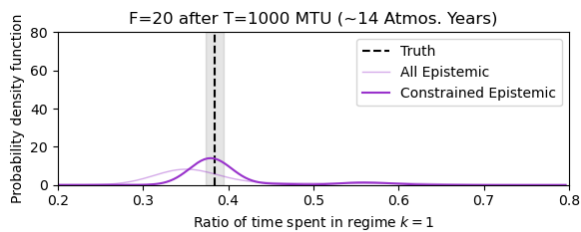

b)
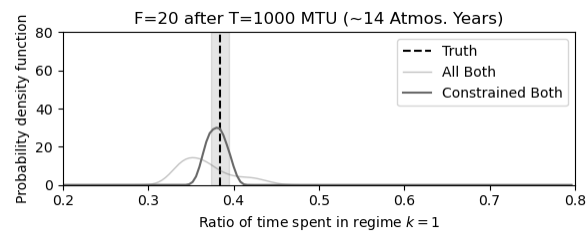

c)
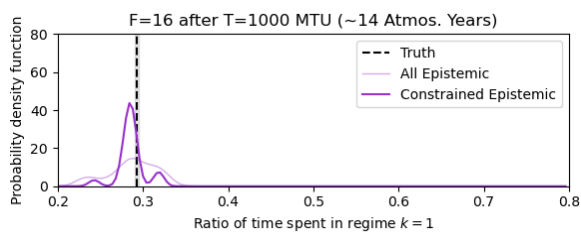

d)
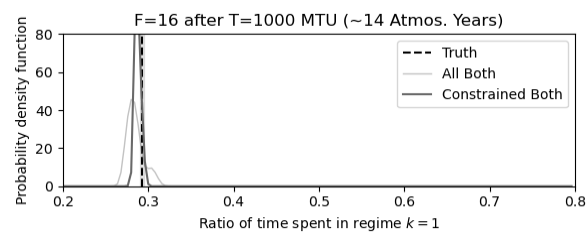

e)
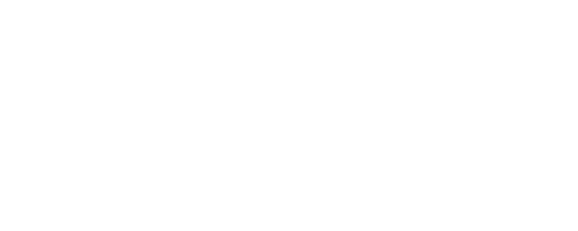

f)
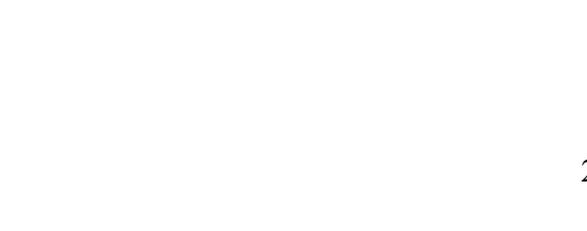

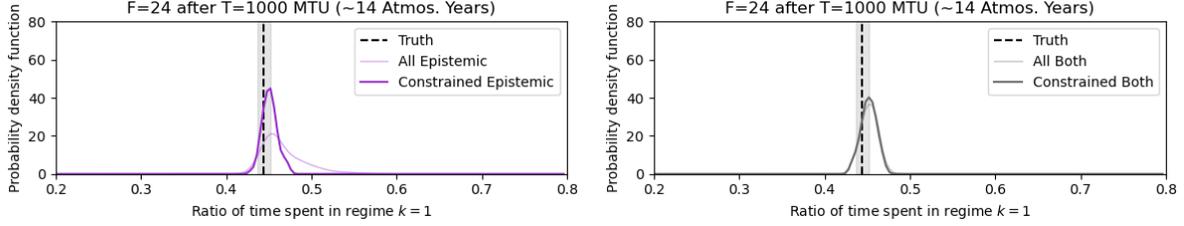

Fig. 10 Distributions showing ratio of time spent in regime $k = 1$ across all ensemble members (light) and constrained ensemble members (bold) at T=1000 MTU, for (a-b) $F = 20$, (c-d) $F = 16$ and (e-f) $F = 24$, where (a,c,e) shows the epistemic ensembles and (b,d,f) the ensembles that sample both uncertainty types. The constrained ensembles are selected as ensemble members that agree with the truth to within 1 standard deviation after T=200MTU in the $F = 20$ simulations (grey shading in Fig. 9a).

This simple approach highlights how we could potentially constrain parameter values based on observations and reduce model uncertainty. The constrained ensemble size could be reduced further, for instance, by constraining over longer periods of time or with stricter targets. This type of approach draws parallels with "History Matching", where PPEs are used to identify regions of the parameter space that agree best with observations in the past (Williamson et al., 2013). History matching takes this further by using emulators that predict a target variable given the parameter values, allowing us to fully probe the parameter space and to predict another "wave" of parameter values for the next PPE. The tuning is repeatedly carried out until reaching sufficient accuracy or until exhausting computational resources. History matching and other calibration techniques are usually designed for problems where there are $O(10)$ parameters. However, neural network parameterisations typically have $O(10^5 - 10^7)$ parameters. There is a need for further exploration of advanced calibration methods that constrain neural network parameters once coupled online.

## 6. Discussion and Conclusions

Using a Bayesian Neural Network, we have shown how we can identify epistemic (model) and aleatoric (data) uncertainties in parameterisations, including modelling aleatoric uncertainty as a function of the input data (heteroscedasticity). On short, weather timescales, the dominant source of uncertainty is aleatoric uncertainty which arises from the subgrid variability. This validates stochastic machine learning approaches that capture this form of uncertainty by training on probabilistic loss functions (e.g., Behrens et al., 2025; Guillaumin & Zanna, 2021). On longer, climate timescales, however, it is crucial to consider epistemic uncertainty. This is because including subgrid variability mostly influences short-term fluctuations and should not affect long-term climate statistics, whereas parameter choices remain fixed within a model and can alter the simulated climate. We find that model uncertainty remains approximately constant across long timescales, in agreement with previous GCM studies (Hawkins & Sutton, 2009). Furthermore, neglecting parametric uncertainty can lead to overconfident responses to changing forcings and should therefore be accounted for in climate change simulations (e.g., Forster et al., 2013; Murphy et al., 2004).

We find that the approach to sampling uncertainty is also an important consideration when building stochastic parameterisations. For aleatoric uncertainty, including temporal correlations on timescales associated with the subgrid variability, for instance through an AR1 process, is essential for producing well-calibrated ensembles in which spread and error are correlated (Arnold et al., 2013). More reliable ensembles can also be achieved by representing aleatoric uncertainty as input dependent (heteroscedasticity), which can be learned by the neural network. For epistemic uncertainty, keeping parameters fixed throughout the simulation better represents the constant parametric uncertainty we aim to capture (Hawkins & Sutton, 2009).

We also explored how PPEs can be used to fine-tune parameter values. While PPEs traditionally focus on $O(10)$ parameters, we extend this to a small neural network with ~500 parameters. We showed how the parameters can be crudely constrained based on observations and how this led to improved mean behaviour and reduced model uncertainty. This also held up under changing forcing experiments. Importantly, we found that including aleatoric uncertainty through the stochastic parameterisation improved the success of parameter calibration. If this behaviour holds in full GCMs, it could prove valuable for both physics-based and ML parameterisations, particularly given the potential for compensating errors that can arise during calibration – for example, from the interaction between resolved and unresolved atmospheric gravity waves (Cohen et al., 2013; Mansfield & Sheshadri, 2022) or from competing radiative effects of clouds (Ma et al., 2022; Zhao et al., 2022). These challenges, often referred to as 'over-tuning', can limit the broader adoption of calibration techniques (Hourdin et al., 2017; Williamson et al., 2017). To extend this research to full GCMs, more advanced calibration should be explored, such as history matching, ensemble Kalman methods, or Bayesian optimisation (Dunbar et al., 2021; King et al., 2024; Watson-Parris et al., 2021; Williamson et al., 2013). There is also the question of how to scale up PPE to neural network parameterisations with $O(10^5 - 10^7)$. For this, it may be worth considering sampling approaches to improve computational efficiency and reduce redundancy in the ensemble, for instance by sampling the full parameter space or increasing parameter diversity (Karmalkar et al., 2019; Sexton et al., 2019, 2021).

This work provides insights into which types of uncertainty to target and how best to sample them across timescales. However, we used a simplified dynamical system as a toy model of the Earth's atmosphere. The next step is to explore how well this holds up in a more realistic GCM. In doing so, we expect that computational cost could become a bottleneck, since BNNs require frequent sampling of high dimensional distributions and running large member ensembles to capture online uncertainties. Further exploration of cheaper approaches to capturing epistemic and aleatoric uncertainties may be required. The approaches outlined here serve as a starting point and we hope that this contributes to a growing emphasis on uncertainty quantification for ML parameterisations.

Although aleatoric uncertainty alone may suffice for weather prediction, our results suggest that models designed for seamless prediction across timescales (such as the Met Office Unified Model, Brown et al., 2012) must sample both aleatoric and epistemic uncertainty. These results are obtained in a highly idealised system, yet they highlight fundamental mechanisms likely relevant to more complex models. If similar behaviour holds in full GCMs, then both sources of uncertainty should be incorporated into parameterisation development intended for use across timescales. More generally, Earth system prediction (whether fully data-driven models, hybrid GCMs, or entirely physics-based) should represent both aleatoric and epistemic uncertainty to capture the full range of possible outcomes across weather and climate regimes.

*Acknowledgments.*

We are grateful to Yee Whye Teh for the insightful discussions. This research received support through Schmidt Sciences, LLC. HMC was supported by the EERIE project (Grant Agreement No 101081383) funded by the European Union. Views and opinions expressed are however those of the author(s) only and do not necessarily reflect those of the European Union or the European Climate Infrastructure and Environment Executive Agency (CINEA). Neither the European Union nor the granting authority can be held responsible for them. University of Oxford's contribution to EERIE is funded by UK Research and Innovation (UKRI) under the UK government's Horizon Europe funding guarantee (grant number 10049639). HMC was also supported by the Leverhulme Trust Research Project Grant `Exposing the nature of model error in weather and climate models' and through a Leverhulme Trust Research Leadership Award.

*Data Availability Statement.*

All code used to produce these simulations and all plots is available at https://github.com/lm2612/L96_UQ.

# Supplementary Material

**Contents**

- **Supplementary Text 1: Bayesian Neural Networks**
- **Supplementary Text 2: AR1 Implementation**
- **Supplementary Figures**

**Supplementary Text 1. Bayesian Neural Networks**

Bayesian Neural Networks (BNNs) assume that the parameters of the neural networks are random variables, $\theta$, with associated probability distributions. As with any Bayesian technique, we start by defining prior distributions over these model parameters, $p(\theta)$. Training BNNs requires carrying out Bayesian inference to condition on observed data, $D$, and learn the posterior distribution as

$$p(\theta|D) = \frac{p(D|\theta)p(\theta)}{\int_{\theta'} p(D|\theta')p(\theta')d\theta'}$$

Equation S6

This is Bayes' Theorem. Exact Bayesian inference is intractable for neural networks because it requires integrating over all possible values of $\theta$ (the denominator in Equation S6). While the posterior distribution can be estimated numerically using Markov chain Monte Carlo (MCMC), this relies on sampling which is **computationally expensive** and often **slow to converge**. This is especially true when $\theta$ is high-dimensional, as is the case with deep neural networks. Instead, we can leverage an approximate Bayesian inference method called **variational inference (VI)**, which approximates the true posterior with a simpler distribution, known as the **variational distribution** or **guide** (Blei et al., 2017). VI becomes an optimization problem, where the goal is to minimize the distance between the variational distribution, $q(\theta)$, and the true posterior, $p(\theta|D)$, and removes the need for sampling. To measure the closeness between two distributions, we use the Kullback–Leibler (KL) divergence:

$$KL(q(\theta)||p(\theta,D) = \int_q q(\theta)\log\left(\frac{q(\theta)}{p(\theta|D)}\right) = E_{q(\theta)}[\log\left(\frac{q(\theta)}{p(\theta|D)}\right)]$$

$$KL(q(\theta)||p(\theta,D) = E_{q(\theta)}(\log(q(\theta))) - E_{q(\theta)}(\log p(\theta|D))$$

Variational inferences seeks $q(\theta)$ which minimizes this distance:

$$q(\theta)^* = \underset{q(\theta)}{\mathrm{argmin}}(KL(q(\theta)||p(\theta,D))$$

If we write the posterior in terms out in full,

$$KL(q(\theta)||p(\theta,D) = E_{q(\theta)}(\log(q(\theta))) - E_{q(\theta)}(\log p(\theta,D)) + \log(D)$$

we notice there is a log evidence term, $\log(D)$, which is generally intractable. We instead optimise a different objective consisting of first two terms

$$ELBO = E_{q(\theta)}(\log p(\theta,D)) - E_{q(\theta)}(\log(q(\theta)))$$

So we can write KL divergence as

$$KL(q(\theta)||p(\theta,D) = -ELBO + \log(D)$$

Note that since $KL(q(\theta)||p(\theta,D)$, $\log(D) > ELBO$, hence the name evidence lower bound. By maximising the ELBO we are minimising the KL divergence.

The ELBO is estimated using (Ranganath et al., 2013; Wingate & Weber, 2013). Then, we can leverage standard optimisation techniques, such as stochastic gradient descent, to optimise the variational distribution. This scales well to high dimensions and large datasets. One of the main challenges comes with choosing an appropriate variational distribution that is flexible enough to approximate the true posterior but simple enough for efficient optimisation (Blei et al., 2017). A common choice in BNNs the Gaussian distribution, which we will use here (Barber & Bishop, 1997; Hinton & van Camp, 1993). Although it is common to restrict the covariance matrix across all weights to being diagonal, in other words, all weights are independent of each other (mean field approach), we will use a multivariate distribution, allowing for correlations between weights for increased flexibility (Barber & Bishop, 1997; Hinton & van Camp, 1993).

This means

$$q(\theta) = \mathcal{N}_p(\mu_\theta, \Sigma_\theta)$$

where $p$ is the number of neural network parameters. In variational inference, we learn the mean of the multivariate Gaussian, $\mu_\theta$, a vector of length $p$, and the covariance matrix, $\Sigma_\theta$, a covariance matrix of size $p \times p$ (although note the upper and lower triangles are the same). There are $\left(\frac{p^2}{2} + p\right)$ parameters that define the distribution $q(\theta)$.

**Supplementary Text 2. AR1 Implementation**

**AR1 Process for Stationary Timeseries**

Wilks (2006) defines the model for a **stationary continuous time series**

$$y_t = \mu + \phi(y_{t-1} - \mu) + \epsilon_t$$

where $\mu$ is the mean, $\phi$ is the autoregressive parameter, and $\epsilon_t$ is a noise term, known as the "innovation". The first terms are deterministic ($\mu + \phi(y_{t-1} - \mu)$) and the last term, $\epsilon_t$, adds a random shock to the system. We assume $\epsilon_t$ are independent and Gaussian distributed with mean 0 and variance $\sigma_\epsilon^2$, known as the residual variance.

If we assume the data to be Gaussian distributed with mean, $\mu$, and variance, $\sigma_y^2$, we can compute the residual variance from the timeseries as

$$\sigma_\epsilon^2 = Var(y_t) - Var(\phi * y_{t-1})) = \sigma_y^2 - \phi^2 \sigma_y^2 = (1 - \phi^2)\sigma_y^2$$

$$\sigma_\epsilon^2 = (1 - \phi^2)\sigma_y^2$$

where $\phi$ is the autoregressive parameter and is equal to the lag-1 autocorrelation of the timeseries.

We can write

$$y_t = \mu + \phi \epsilon_{t-1} + \epsilon_t$$

where

$$\epsilon_{t-1} = \mu - y_{t-1}$$

and

$$\epsilon_t \sim \mathcal{N}(0, (1 - \phi^2)\sigma_y^2)$$

**AR1 Process for Aleatoric Uncertainty**

Now, we consider our case where our data is non-stationary, and we want to learn the AR1 process that captures aleatoric uncertainty alone. Instead of using the mean, $\mu$, we will use, $y = f_{\bar{\theta}}(x)$, a deterministic function where the parameters are fixed at their median values,

$$y_t = f_{\bar{\theta}}(x_t) + \phi \epsilon_{t-1} + \epsilon_t$$

where the residual from the previous timestep becomes

$$\epsilon_{t-1} = y_{t-1} - f_{\bar{\theta}}(x_{t-1})$$

and the innovation is the same as above,

$$\epsilon_t \sim \mathcal{N}(0, (1-\phi^2)\sigma_y^2)$$

where $\sigma_y^2$ is now aleatoric uncertainty we learned in the model during training. For the standard BNN, this is just a fixed value for all values of x, but for the heteroscedastic BNN, this is obtained from a pass through the network.

This gives

$$\epsilon_t \sim (1-\phi)^{\frac{1}{2}} \sigma_y \, \mathcal{N}(0,1)$$

where $\Sigma(0,1)$ is a random number drawn from a Gaussian

**AR1 Process for Epistemic Uncertainty**

For the epistemic uncertainty, our function depends on $x_t$ and also parameters, $\theta$,

$$y_t = f(x_t|\theta)$$

where the parameters are sampled from the variational distribution. Here this is a multivariate Gaussian distributed with size $p$:

$$\theta \sim \mathcal{N}_p(\mu_\theta, \Sigma_\theta)$$

We also have a deterministic function, $f_{\bar{\theta}}(x_t)$. We use almost the same formulation as above,

$$y_t = f_{\bar{\theta}}(x_t) + \phi \epsilon_{t-1} + \epsilon_t$$

where

$$\epsilon_t \sim \mathcal{N}\left(0, (1-\phi^2)\sigma_y^2\right)$$

The main difference lies in how we estimate $\sigma_y^2$. When considering epistemic uncertainty, this is the variance of the function evaluated at $x_t$ *after integrating over all possible parameter values*, $\theta$,

$$\sigma_y^2 = \int_\theta \text{Var}\left(f(x_t|\theta)\right) d\theta = \int_\theta (f(x|\theta_i) - f_{\bar{\theta}}(x))^2 \, p(\theta) d\theta$$

which is

$$\sigma_y^2 = \sum_{\theta_i \sim \mathcal{N}_p(\mu_\theta, \Sigma_\theta)}^{\theta_N} (f(x|\theta_i) - f_{\bar{\theta}}(x))^2$$

where the samples $\theta_i$ from $i = 1, \ldots N$ are drawn from the learned multivariate normal distribution $\mathcal{N}_p(\mu_\theta, \Sigma_\theta)$. This gives

$$\epsilon_t \sim N\left(0, (1-\phi^2) \sum_{\theta_i \sim \mathcal{N}_p(\mu_\theta, \Sigma_\theta)}^{\theta_N} \left(f(x|\theta_i) - f_{\bar{\theta}}(x)\right)^2\right)$$

$$\epsilon_t \sim \left[(1-\phi^2) \sum_{\theta_i \sim \mathcal{N}_p(\mu_\theta, \Sigma_\theta)}^{\theta_N} \left(f(x|\theta_i) - f_{\bar{\theta}}(x)\right)^2\right]^{\frac{1}{2}} \mathcal{N}(0,1)$$

**How to decide on $N$?**

Using more samples to approximate this variance should be more accurate but more expensive. In this setting, we use this variance as an estimate of the noise term to be added in the AR1 process. The figure below shows that estimating $\sigma_y^2$ with small $N$ still gives similar ensemble spread throughout the simulation as those that use large $N$. We use $N = 10$ for the rest of the simulations, but expect that $N = 2$ would be sufficient.

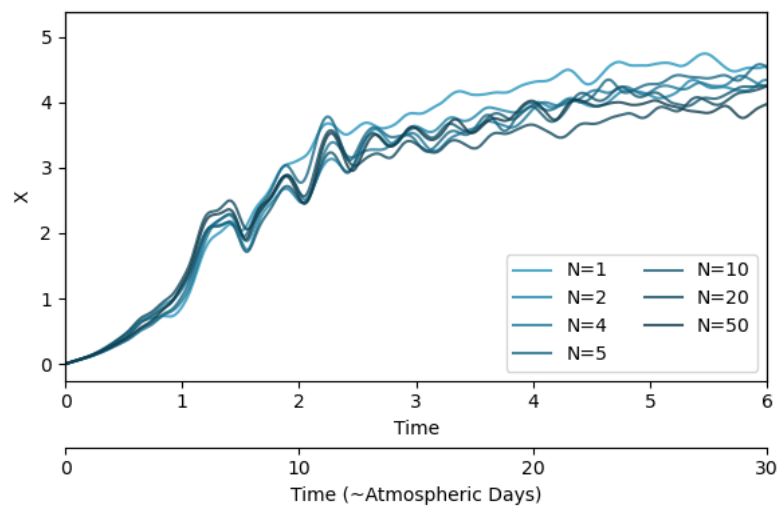

Spread against time for one forecast where we test different $N$ to estimate the epistemic uncertainty.

**Both**

To include both forms of uncertainty, we use the same as above but we sum the variances from the aleatoric component and the epistemic component,

$$\epsilon_t \sim \mathcal{N}\left(0, (1-\phi^2)\left(\sigma_y^2 + \sum_{\theta_i \sim \mathcal{N}_p(\mu_\theta, \Sigma_\theta)}^{\theta_N} \left(f(x|\theta_i) - f_{\bar{\theta}}(x)\right)^2\right)\right)$$

## Supplementary Figures

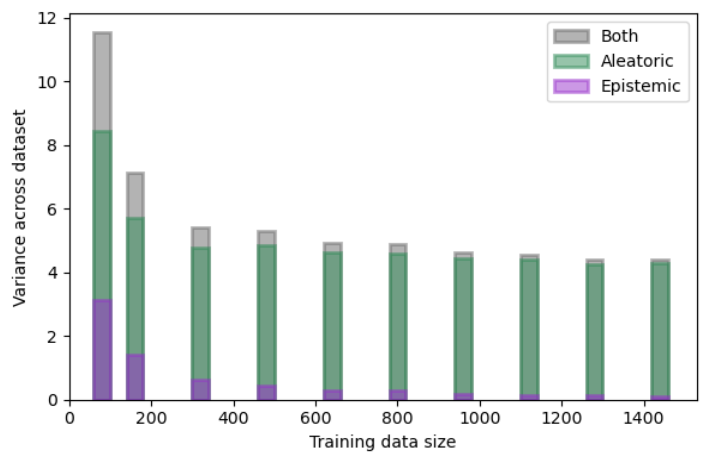

Supplementary Fig. 11 Variance over the dataset for the BNN sampling epistemic, aleatoric and both forms of uncertainty as the training data size is increased. We use 800 training data points.

a) **Independent Noise**     b)     **AR1 Noise**

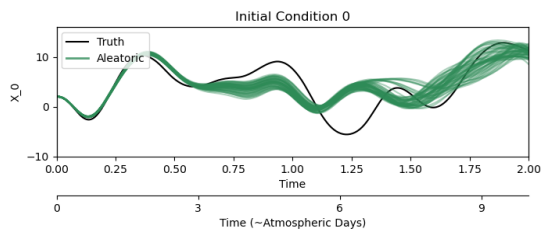
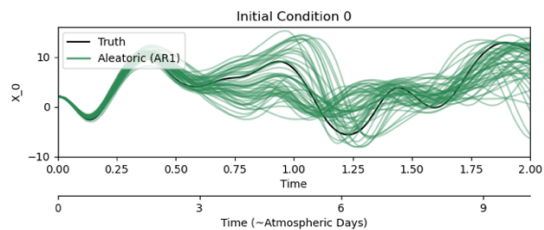

c)     d)

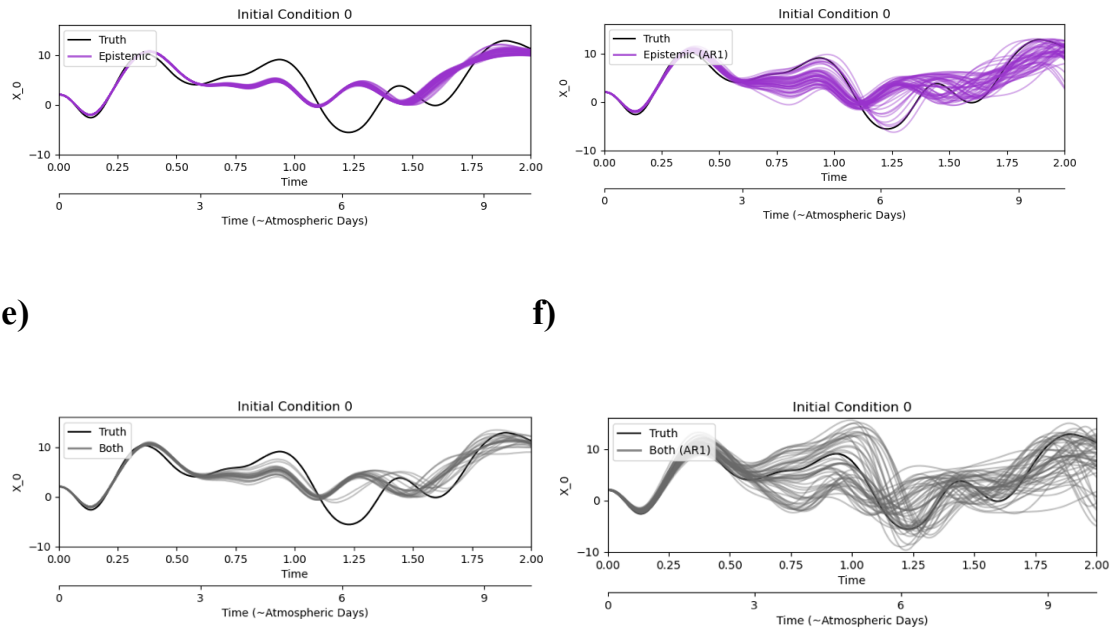

Supplementary Fig. 12. Same as Fig. 5 but for the homoscedastic BNN.

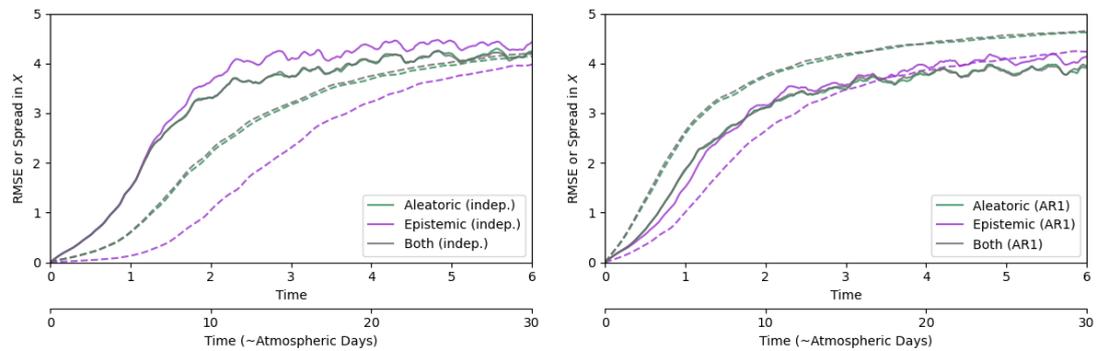

Supplementary Fig. 13. Same as Fig. 6 but for the homoscedastic BNN.